\documentclass{jpsj2}

\title{%
Possible Pairing Symmetry of Three-dimensional Superconductor $\mathrm{UPt_3}$ \\
--- Analysis Based on a Microscopic Calculation ---
}

\author{%
Shogo SHINKAI\thanks{E-mail address: shogo@scphys.kyoto-u.ac.jp} and Kosaku YAMADA
}

\inst{%
Depertment of Physics, Kyoto University, Sakyo-ku, Kyoto 606-8502
}

\recdate{\today}

\abst{%
Stimulated by the anomalous superconducting properties of $\mathrm{UPt_3}$, we investigate the pairing symmetry and the transition temperature in the two-dimensional(2D) and three-dimensional(3D) hexagonal Hubbard model. We solve the $\mathrm{\acute{E}liashberg}$ equation using the third order perturbation theory with respect to the on-site repulsion $U$. As results of the 2D calculation, we obtain distinct two types of stable spin-triplet pairing states. One is the $f$-wave($\mathrm{B_1}$) pairing around $n = 1.2$ and in a small $U$ region, which is caused by the ferromagnetic fluctuation. Then, the other is the $p_x$(or $p_y$)-wave($\mathrm{E_1}$) pairing in large $U$ region far from the half-filling ($n = 1$) which is caused by the vertex corrections only. However, we find that the former $f$-wave pairing is destroyed by introduced 3D dispersion. This is because the 3D dispersion breaks the favorable structures for the $f$-wave pairing such as the van Hove singularities and the small pocket structures. Thus, we conclude that the ferromagnetic fluctuation mediated spin-triplet state can not explain the superconductivity of $\mathrm{UPt_3}$. We also study the case of the pairing symmetry with a polar gap. This $p_z$-wave($\mathrm{A_1}$) is stabilized by the large hopping integral along c-axis $t_z$. It is nearly degenerate with the suppressed $p_x$(or $p_y$)-wave($\mathrm{E_1}$) in the best fitting parameter region to $\mathrm{UPt_3}$ ($1.3 \le t_z \le 1.5$). These two $p$-wave pairing states exist in the region far from the half-filling, in which the vertex correction terms play crucial roles like the case in $\mathrm{Sr_2RuO_4}$.
}

\kword{%
$\mathrm{UPt_3}$, third-order perturbation theory, spin-triplet superconductivity, heavy fermion
}

\begin{document}
\maketitle

\section{Introduction}
Much attention have been focused on hexagonal heavy fermion superconductor $\mathrm{UPt_3}$, since it was discovered in 1984\cite{stewart}. The heavy fermion state is characterized by its large value of the electronic specific heat coefficient $\gamma = 420 \mathrm{mJ/K^2 mole U}$. The coefficient of $T^2$-law of resistivity is also enhanced by $10^5$ times as large as that of conventional metal. Such behavior arises from the strong electron correlation. In $\mathrm{UPt_3}$, the itinerant electrons compose such a Fermi liquid state and then undergo superconducting transition. In zero field, the two superconducting transitions occur at $\sim$0.48 K and $\sim$0.52 K \cite{fisher}\cite{hassel}. The field-temperature phase diagram was constructed experimentally \cite{adenwalla} and it consists of at least three phases (denoted as A, B and C). Most of theoretical models for this splitting of $T_{\mathrm{c}}$ are based on the coupling between the superconducting order parameter and the small antiferromagnetic moment ($\mu_S = 0.02 \mu_B/\mathrm{U}$) which appears below $T_{\mathrm{N}} \sim 5 \mathrm{K}$ \cite{aeppli}\cite{heffner}. 

Furthermore, a power law behavior of the ultrasound attenuation \cite{Shivaram}, NMR relaxation rate \cite{kohori}, specific heat \cite{fisher}\cite{hassel} and penetration depth \cite{broholm} below $T_{\mathrm{c}}$ suggested an unusual order parameter with energy gap which vanishes at some points on the Fermi surface. In addition, the NMR \cite{kohori} \cite{tou} and muon spin rotation-relaxation ($\mathrm{\mu SR}$) \cite{luke} experiments detected no change of the Knight shift below $T_{\mathrm{c}}$. These suggest a possibility of the odd-parity of the Cooper pairing.

In order to explain these unconventional properties in $\mathrm{UPt_3}$, various phenomenological theories have been proposed. Theories based on two-dimensional representation\cite{joynt}\cite{sauls} assign a gap with both point nodes along c-axis and a line of nodes at $k_z = 0$ (called a hybrid gap) to the phase B. On the other hand, the theory based on one-dimensional representation\cite{machida} requires either a hybrid gap or a gap with only a line of nodes at $k_z = 0$ (a polar gap). The experiments of thermal conductivity\cite{lussier1}\cite{lussier2} and penetration depth\cite{broholm} suggested that a hybrid gap is favored, but the gap structure of $\mathrm{UPt_3}$ has not been completely determined.

In contrast with the phenomenological theories, few microscopical theories determining the pairing symmetry and the gap structure have been reported. The band calculations\cite{oguchi}\cite{norman}\cite{kimura1}, based on the assumption that 5f-electrons are itinerant, reported that five three-dimensional bands form the Fermi surface. The results of the de Haas-van Alphen (dHvA) experiments \cite{taillefer}\cite{kimura1}\cite{kimura2} are consistent with these band calculations, and the observed effective mass of each band is almost 20 times as large as the corresponding calculated mass. Among these five bands, we take up band 37-electron in ref.19 (a rugby ball like band structure with small electron pockets at K-points in the hexagonal Brillouin zone) because it has the heaviest effective mass and occupies the largest area of the Fermi surface. Then, in the following sections, we discuss the pairing symmetry in the microscopic way based on band 37-electron.

\section{Model}
We consider the following single band Hubbard model on hexagonal lattice, 
\begin{align}
H &= H_0 + H_{\mathrm{int}} , \\
H_0 &= \displaystyle \sum_{\vec{k}\sigma} \xi(\vec{k}) c^{\dagger}_{\vec{k}\sigma} c_{\vec{k}\sigma} , \\
H_{\mathrm{int}} &= \frac{U}{2N} \sum_{\vec{k}_i} \sum_{\sigma \neq \sigma'} \delta_{\vec{k}_1+\vec{k}_2 , \vec{k}_3+\vec{k}_4} c^{\dagger}_{\vec{k}_1\sigma} c^{\dagger}_{\vec{k}_2\sigma'} c_{\vec{k}_3\sigma'} c_{\vec{k}_4\sigma} .
\end{align}
From the tight binding approximation, energy dispersion $\xi(\vec{k})$ is given by
\begin{align}
\xi(\vec{k}) = & -2 t_1 \Bigl( 2 \cos {\frac{\sqrt{3} k_x}{2}} \cos {\frac{k_y}{2}} + \cos {k_y} \Bigr) \nonumber \\
& -2 t_2 \Bigl( \cos {\sqrt{3} k_x} + 2 \cos {\frac{\sqrt{3} k_x}{2}} \cos {\frac{3 k_y}{2}} \Bigr) \nonumber \\
& -2 t_3 \Bigl( 2 \cos {\sqrt{3} k_x} \cos {k_y} + \cos {2 k_y} \Bigr) \nonumber \\
& -2 t_z \Bigl( \cos {k_z} \Bigr) - \mu .
\end{align} 
Here, $t_1$, $t_2$ and $t_3$ are the nearest-, the next-nearest-, and the third-nearest-neighbor hopping integrals in ab-plane respectively. The $t_z$ is the hopping integral along c-axis, and $\mu$ is the chemical potential. Now we fix $t_1 = 1$, and calculate physical quantities changing the other variables $t_2$, $t_3$, $t_z$, $U$ and electron filling $n$.

\section{Formulation}
One of the authors (K.Y.) has described the Fermi liquid theory for the heavy electron system \cite{yamada}\cite{nisikawa}. Based on this idea, we start from the renormalized quasiparticle state, and calculate the momentum dependence of the effective interaction between them. 

Recently, some authors have calculated the transition temperture $T_{\mathrm{c}}$ and the pairing symmetry for many superconductors including high-$T_{\mathrm{c}}$ cuprates and $\mathrm{Sr_2RuO_4}$ on the basis of the third order perturbation theory (TOPT) with respect to $U$ \cite{yanase}. All of them are in good agreement with experimental facts. Although the convergence of the vertex correction terms is an important problem, TOPT in the moderately correlated region has been justified by higher order calculations \cite{nomura}. 
Thus, in the same framework, we apply TOPT to our model.

In TOPT, the effective interaction for the spin-singlet state is given by
\begin{align}
V^{\mathrm{S}}(k,k') &= V^{\mathrm{S}}_{\mathrm{RPA}}(k,k') + V^{\mathrm{S}}_{\mathrm{Vertex}}(k,k') \\
V^{\mathrm{S}}_{\mathrm{RPA}}(k,k') &= U + U^2 \chi_0(k-k') + 2 U^3 \chi_0^2(k-k') \\
V^{\mathrm{S}}_{\mathrm{Vertex}}(k,k') &= 2 U^3 \mathrm{Re} \Bigl[ \sum_q G(k+q) G(k'+q) [\chi_0(q) - \phi_0(q)] \Bigr].
\end{align}
For the spin-triplet state,
\begin{align}
V^{\mathrm{T}}(k,k') &= V^{\mathrm{T}}_{\mathrm{RPA}}(k,k') + V^{\mathrm{T}}_{\mathrm{Vertex}}(k,k') \\
V^{\mathrm{T}}_{\mathrm{RPA}}(k,k') &= - U^2 \chi_0(k-k') \\
V^{\mathrm{T}}_{\mathrm{Vertex}}(k,k') &= 2 U^3 \mathrm{Re} \Bigl[ \sum_q G(k+q) G(k'+q) [\chi_0(q) + \phi_0(q)] \Bigr],
\end{align}
where
\begin{align}
\chi_0(q) &= - \sum_k G(k+q) G(k)\\
\phi_0(q) &= - \sum_k G(q-k) G(k)\\
G(k) &= \frac{1}{i\omega_n - \xi(\vec{k})}.
\end{align}
Here, $k$ is a shorthand notation as $k = (\vec{k},i\omega_n)$ and $\omega_n = \pi T (2n+1)$ is a fermion Matsubara frequency. $\xi(\vec{k})$ is the band of quasiparticles with large effective mass. In Fig.1, the diagrams of the effective interaction are shown. The diagrams enclosed by a dashed line are the vertex correction terms. We calculate $T_{\mathrm{c}}$ and the pairing symmetry by solving the $\mathrm{\acute{E}liashberg}$ equation,
\begin{align} 
\lambda_{\mathrm{max}} \Delta(k) &= - \frac{T}{N}\sum_{k'} V(k,k')|G(k')|^2 \Delta(k').
\end{align}
At $T=T_{\mathrm{c}}$, the maximum eigenvalue $\lambda_{\mathrm{max}}$ becomes unity, and the pairing symmetry is determined by the momentum dependence of the gap function $\Delta(k)$. Thus, by estimating $\lambda_{\mathrm{max}}$, we can determine which type of pairing symmetry is stable. Here, we have ignored the damping effect due to the normal selfenergy. This contribution is important for estimating $T_{\mathrm{c}}$ quantitatively, but qualitative properties of superconductivity such as the pairing symmetry is not usually influenced.

\begin{figure}[t]
\begin{center}
\includegraphics[width=8cm]{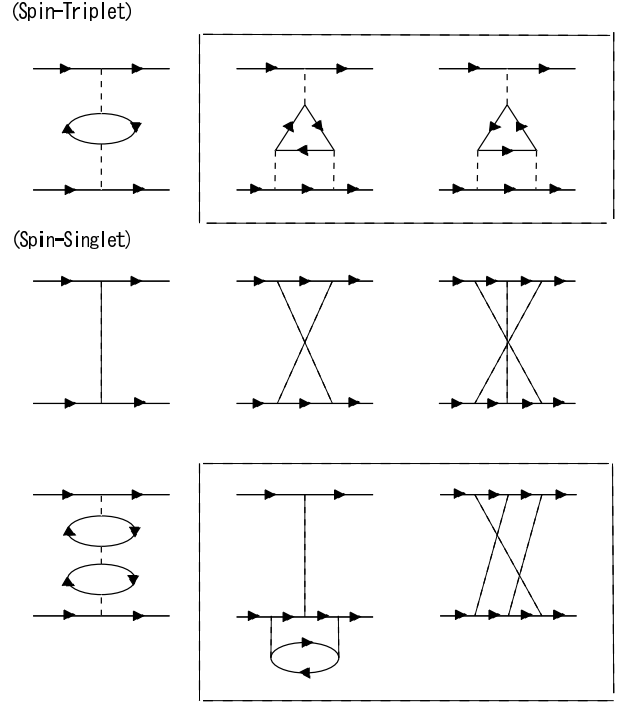}
\end{center}
\caption{Feynman diagrams of the effective interaction up to third order. Solid and dashed lines correspond to the Green's function and the interaction, respectively.}
\label{f1}
\end{figure}

\section{Results of 2D Calculations}
In this section, we show the results of two-dimensional calculations, considering the $k_z = 0$ plane (including $\mathrm{\Gamma}$, K, M points in the hexagonal Brillouin zone) of the band 37, that is, we fix $t_z = 0$. Here, we notice that band 37 is highly three-dimensional, but the calculations on the basal triangular lattice gives us much information as starting points. We will show the results of 3D calculations by introducing $t_z$ in \S 5. Then, we also fix $t_3 = 0$ and $T = 0.01$ in this section. The calculated Fermi surfaces are shown in Fig.2. We can see larger electron pockets at K-points as either $n$ or $t_2$ increases. Within the $k_z = 0$ plane, the best fitting parameter region for $\mathrm{UPt_3}$ is $t_2 = 0.4 \sim 0.6, n = 0.7 \sim 1.0$. The possible candidate of pairing symmetry in 2D triangular lattice belongs to the following irreducible representations of $C_{6v}$.
\begin{tabular}{ll}
\hline
symmetry   & One of basis function  \\
\hline
$\mathrm{B_1}$($f_1$-wave)  & $\sin{\frac{k_y}{2}} \Bigl(\cos{\frac{\sqrt{3} k_x}{2}}-\cos{\frac{k_y}{2}} \Bigr)$  \\
$\mathrm{B_2}$($f_2$-wave)  & $\sin{\frac{\sqrt{3} k_x}{2}} \Bigl(\cos{\frac{\sqrt{3} k_x}{2}}-\cos{\frac{3k_y}{2}} \Bigr)$  \\
$\mathrm{E_1}$($p$-wave)  & $\left\{ 
\begin{array}{ll}
\sin{\frac{\sqrt{3} k_x}{2}} \cos{\frac{k_y}{2}}  \\
\cos{\frac{\sqrt{3} k_x}{2}} \sin{\frac{k_y}{2}} + \sin{k_y}  
\end{array}
\right. $  \\
$\mathrm{E_2}$($d$-wave)  & $\left\{
\begin{array}{ll}
\sin{\frac{\sqrt{3} k_x}{2}} \sin{\frac{k_y}{2}}  \\
\cos{\frac{\sqrt{3} k_x}{2}} \cos{\frac{k_y}{2}} -\cos{k_y}
\end{array}
\right. $  \\
\hline
\end{tabular}

\begin{figure}[t]
\begin{center}
\includegraphics[width=8cm]{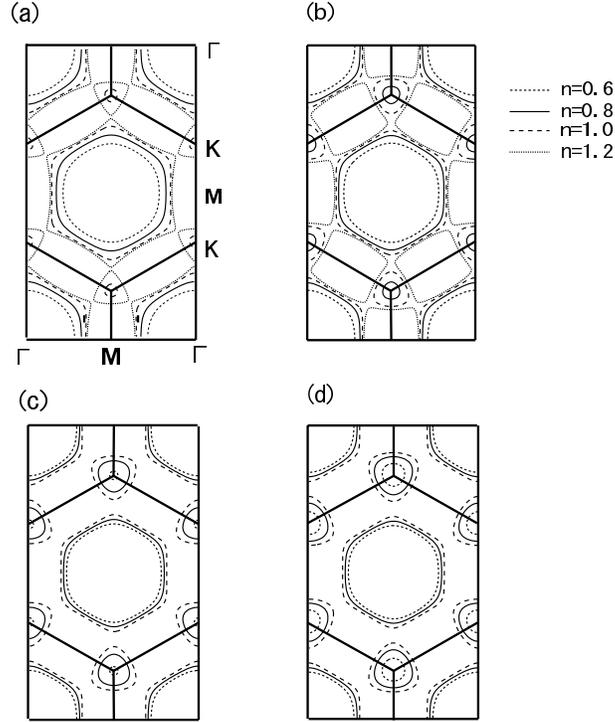}
\end{center}
\caption{Calculated Fermi surfaces in the case of (a)$t_2 = 0.3$, $n = 0.6, 0.8, 1.0, 1.2$ (b)$t_2 = 0.4$, $n = 0.6, 0.8, 1.0, 1.2$ (c)$t_2 = 0.5$, $n = 0.6, 0.8, 1.0$ (d)$t_2 = 0.6$, $n = 0.6, 0.8, 1.0$ . Electron pockets at K-points become large as either $n$ or $t_2$ increases.}
\label{f2}
\end{figure}

First, we set $U = 2$. Figure 3 is a contour plot of $\lambda_{\mathrm{max}}$ as a function of $t_2$ and electron filling $n$ ($n = 1$ at the half filling). The black thick line represents the crossover line of two pairing symmetries. When $n$ is larger than half filling, an $f_1(\mathrm{B_{1}})$-wave spin-triplet state is stable. This $f_1$-wave state was predicted by Ikeda $et\ al.$\cite{ikeda} in the case of $\mathrm{Na_xCoO_2 \cdot yH_2O}$. In contrast, for the case less than half filling, $d(\mathrm{E_{2}})$-wave pairing state covers a wide region. But the eigenvalues of this $d$-wave state is too low to be realized. Then, what is the origin of $f$-wave pairing above half filling? In figure 4, we show the bare spin susceptibility $\chi_0(\vec{q},0)$ in the case of $t_2 = 0.3, n = 1.2$. The peak structures can be seen at $\mathrm{\Gamma}$-points, and cannot be seen in the region where $d$-wave pairing state is stable. This peak corresponds to the increase of the density of states by van Hove singularity, which can be seen around K-point in Fig.2(a) (n=1.2). The peak structure at $\mathrm{\Gamma}$-point implies the ferromagnetic instability. These conditions stabilize the $f_1$-wave triplet pairing. Next, to examine influence of the vertex corrections, we calculate the eigenvalues by including only the RPA-like diagrams of the effective interaction up to the third order. The $n$-dependence of the calculated eigenvalues for $f_1$-wave pairing in the RPA-like calculation, and that for $f_1$- and $d$-wave in TOPT are shown in Fig.5. We can see the vertex correction terms work in favor of realizing the triplet pairing, as Nomura $et \ al.$ predicted in $\mathrm{Sr_2RuO_4}$\cite{nomura}. Here, we have to note that the Fermi surface in this $f_1$-wave region is rather larger than actual one. This does not continue to the best fitting region for $\mathrm{UPt_3}$ ($t_2 = 0.4 \sim 0.6, n = 0.7 \sim 1.0$), because of the absence of the van Hove singularities near the Fermi surface in spite of the existence of the electron pockets. 

\begin{figure}[t]
\begin{center}
\includegraphics[width=8cm]{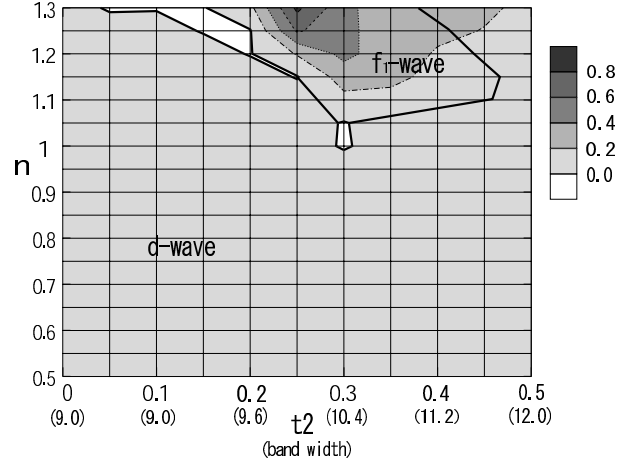}
\end{center}
\caption{The contour plot of $\lambda_{\mathrm{max}}$ as a function of $t_2$ (band width) and $n$ in the case of $U = 2$ and $T = 0.01$. An $f_1$-wave spin-triplet state is stable in the large $n$ region. The eigenvalue for $d$-wave pairing below the half filling is too low to be realized.}
\label{f3}
\end{figure}

\begin{figure}[t]
\begin{center}
\includegraphics[width=8cm]{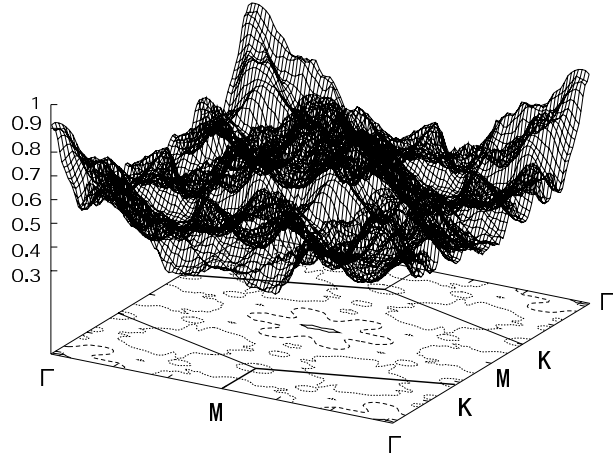}
\end{center}
\caption{The bare spin susceptibility $\chi_0(\vec{q},0)$ in the case of $t_2 = 0.3$, $n = 1.2$ and $T = 0.01$. We can see peak structures around $\mathrm{\Gamma}$-points.}
\label{f4}
\end{figure}

\begin{figure}[t]
\begin{center}
\includegraphics[width=8cm]{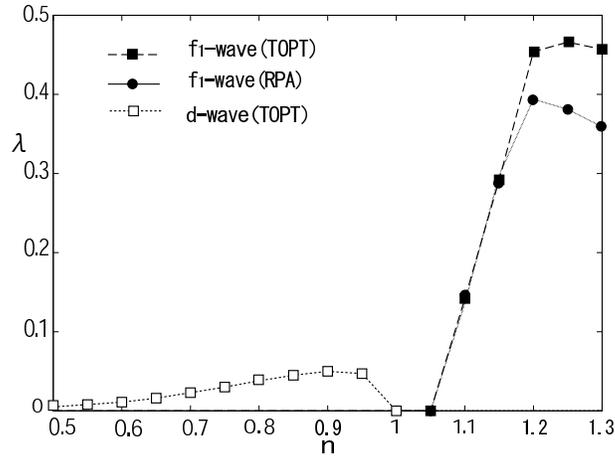}
\end{center}
\caption{The $n$-dependence of eigenvalues in the case of $t_2 = 0.3$, $U = 2$ and $T = 0.01$. The dashed (dotted) line with black (white) squares is the result for $f_1$-wave ($d$-wave) pairing in TOPT. The line with black circles is the result for $f_1$-wave pairing in the RPA-like calculation. }
\label{f5}
\end{figure}

Next, we set $U = 5$ to investigate the triplet pairing in the best fitting parameter region. Figure 6 is a contour plot of $\lambda_{\mathrm{max}}$ as a function of $t_2$ and $n$. We can see $p(\mathrm{E_{1}})$-wave spin-triplet pairing state appears in low $n$ region. In this region, there are no significant peak structures at $\mathrm{\Gamma}$-point, because van Hove singularities and/or small pocket structures are lacking. Thus, this $p$-wave pairing is caused by the vertex correction only, in contrast with the previous $f$-wave pairing. Then we increase $U$ to $U = 7$, and show a contour plot of $\lambda_{\mathrm{max}}$ as a function of $t_2$ and $n$ in Fig.7. We can see the $p$-wave pairing becomes stable. Far from the half-filling, the $p$-wave pairing covers a wide region which includes the best fitting parameter region for $\mathrm{UPt_3}$. Here, we note, in the case of $n \ge 0.9$ for large $U$, we cannot perform reliable calculations because $U$ maybe too large compared with $1/ \chi_0^{\mathrm{max}}$, where $1/ \chi_0^{\mathrm{max}}$ is the maximum value of the bare susceptibility $\chi_0(\vec{q},0)$. 

\begin{figure}[t]
\begin{center}
\includegraphics[width=8cm]{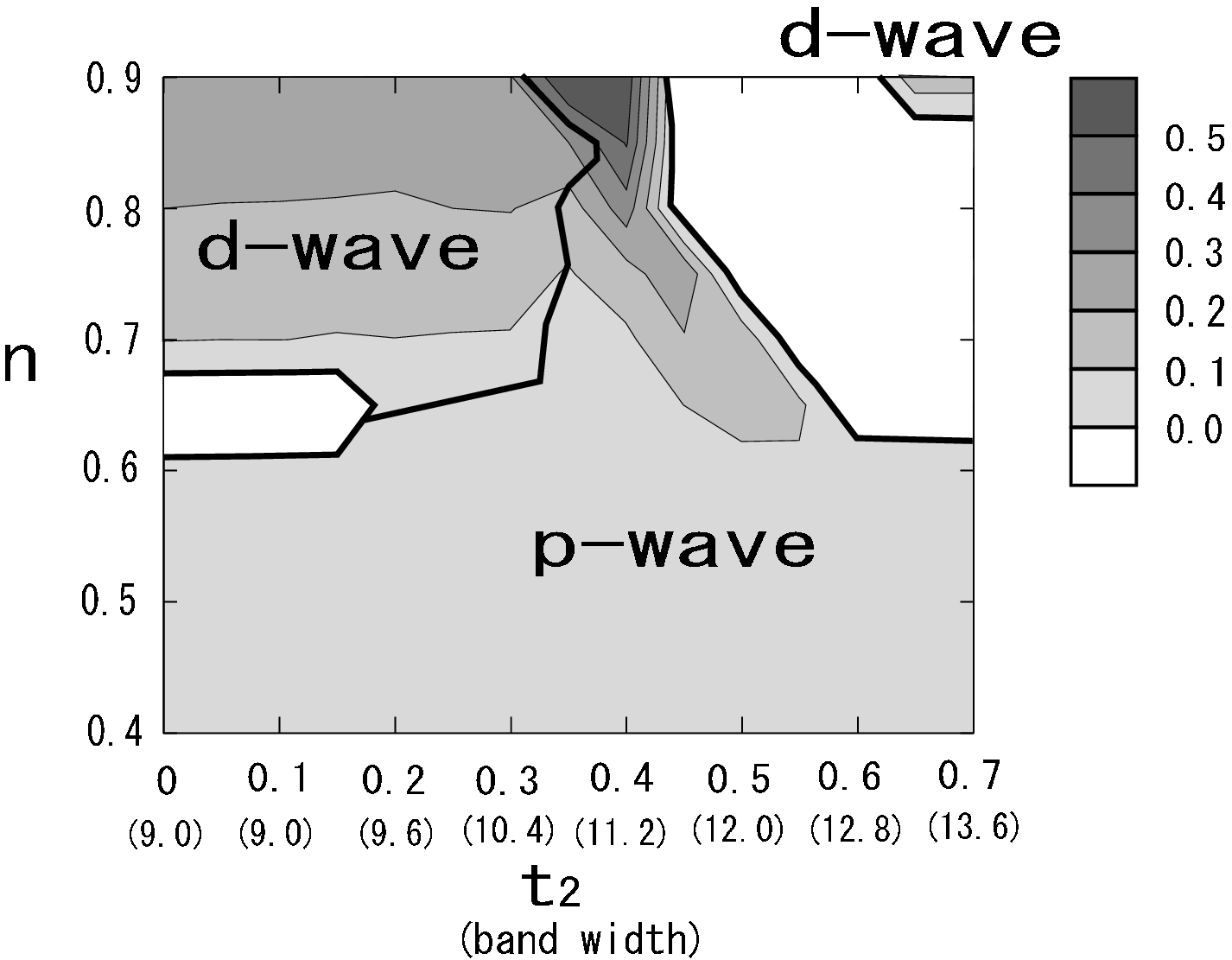}
\end{center}
\caption{The contour plot of $\lambda_{\mathrm{max}}$ as a function of $t_2$ (band width) and $n$ in the case of $U = 5$ and $T = 0.01$. Here, we consider the case of less than half-filling only. A $p$-wave pairing appears far from the half-filling.}
\label{f6}
\end{figure}

\begin{figure}[t]
\begin{center}
\includegraphics[width=8cm]{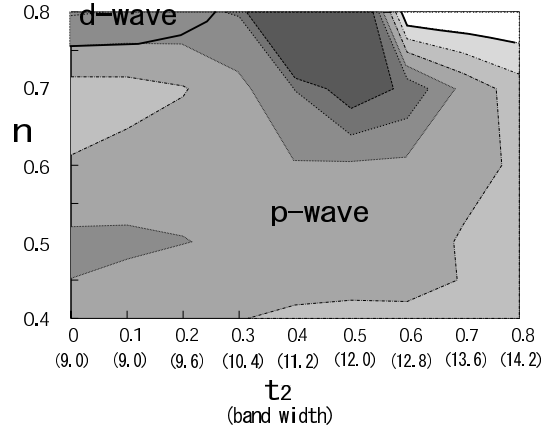}
\end{center}
\caption{The contour plot of $\lambda_{\mathrm{max}}$ as a function of $t_2$ (band width) and $n$ in the case of $U = 7$ and $T = 0.01$. A $p$-wave triplet pairing is stabilized by the vertex corrections for large $U$, and covers the wide region.}
\label{f7}
\end{figure}

In this section, we have obtained distinct two types of stable triplet pairing states. One is the $f(\mathrm{B_{1}})$-wave pairing in large $n$ and small $U$ region, which is caused by the ferromagnetic fluctuation. The vertex correction terms also increase $\lambda_{\mathrm{max}}$ for this $f$-wave pairing. Another is the $p(\mathrm{E_{1}})$-wave pairing in large $U$ region far from the half filling. This state is caused by the vertex corrections only, but the form of the Fermi surface is in better agreement with that from the band calculations. However, since the band 37 is highly three-dimensional, we have to introduce 3D dispersion and investigate the stability for the effect of three-dimensionality.

\section{Results of 3D Calculations}
\subsection{Stability for 3D dispersion}
In this section, we investivate the effect of 3D dispersion on these two pairing states, introducing $t_z$. We also fix $t_3 = 0$ and $T = 0.01$ in this section. First, we apply the 3D calculation to the high-filling $f_1(\mathrm{B_{1}})$-wave pairing state at $U = 2$. In Fig.8, we show the $t_z$-dependence of eigenvalues for $f_1$-wave state in two cases $t_2 = 0.3, n = 1.2$ and $t_2 = 0.3, n = 1.15$. As $t_z$ increases, the eigenvalues show relatively sharp decrease (compared with the next $p$-wave cases) and then vanishes near $t_z = 0.3$. This result is different from the case of $d$-wave pairing on the square lattice\cite{fukazawa}: In that case $T_{\mathrm{c}}$ begins to be suppressed at $t_z = 0.4$, then decreases by one order of magnitude at $t_z = 1.0$. In our case, the rapid decrease of the eigenvalue for $f_1$-wave indicates that the ferromagnetic fluctuation mediated spin-triplet superconductivity is easily destroyed by three-dimensionality, which breaks both the van Hove singularities and the small pocket structures. Actually, the band 37 is highly three-dimensional ($t_z \ge 1$) and both structures exist only near the $k_z = 0$ plane. Thus, we conclude that this $f_1$-wave pairing can not explain the superconductivity of $\mathrm{UPt_3}$.

Next, we apply the 3D dispersion to the low-filling $p(\mathrm{E_{1}})$-wave state at $U = 7$. In Fig.9, we show the $t_z$-dependence of eigenvalues for $p$-wave state in the two cases $t_2 = 0.4, n = 0.6$ and $t_2 = 0.4, n = 0.7$. In contrast to the $f_1$-wave pairing case, the decrease of the eigenvalue of the $p$-wave pairing is rather slow, especially far from the half-filling. This is because the $p$-wave pairing is not caused by the special structures of Fermi surface but the vertex corrections. When we introduce $t_z$, the calculated band width become larger. Thus we increase the on-site repulsion $U$ and try to obtain the reliable eigenvalues in the next subsection.

\begin{figure}[t]
\begin{center}
\includegraphics[width=8cm]{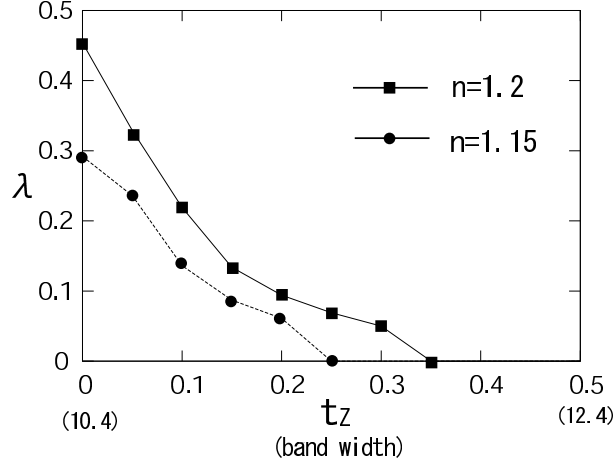}
\end{center}
\caption{The $t_z$-dependence of eigenvalues for $f_1(\mathrm{B_{1}})$-wave pairing in the case of $t_2 = 0.3$, $U = 2$ and $T = 0.01$. The line with black squares (circles) is the result for $n = 1.2$ ($n = 1.15$) in TOPT. The suppression of $\lambda_{\mathrm{max}}$ by 3D dispersion is very strong.}
\label{f8}
\end{figure}

\begin{figure}[t]
\begin{center}
\includegraphics[width=8cm]{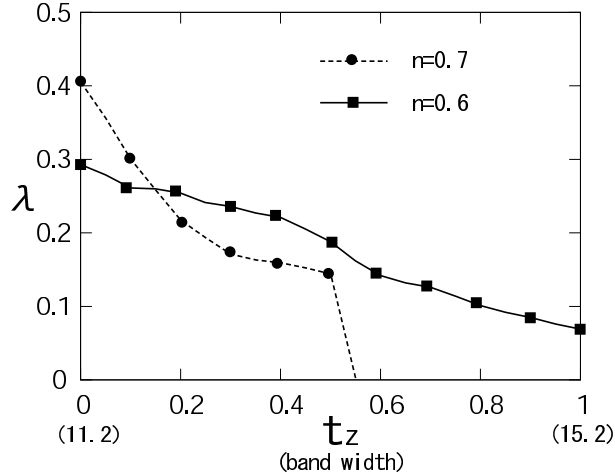}
\end{center}
\caption{The $t_z$-dependence of eigenvalues for $p(\mathrm{E_{1}})$-wave pairing in the case of $t_2 = 0.4$, $U = 7$ and $T = 0.01$. The solid (dashed) line with black squares (circles) is the result for $n = 0.6$ ($n = 0.7$) in TOPT. This low-filling $p$-wave pairing is stable for 3D dispersion compared with the $f_1$-wave in Fig.8.}
\label{f9}
\end{figure}

\subsection{Pairing Symmetry with a Polar Gap}
There are many pairing symmetries in hexagonal lattice ($D_{6h}$) in addition to the irreducible representations of $C_{6v}$, for example $\mathrm{E_{1g}}$($d_{xz}$ and $d_{yz}$ are degenerated), $\mathrm{E_{2g}}$($d_{xy}$ and $d_{x^2-y^2}$ are degenerated), $\mathrm{B_{1g}}$($g$-wave), $\mathrm{B_{2g}}$($g$-wave) for spin-singlet state, $\mathrm{A_{1u}}$($p_z$), $\mathrm{E_{1u}}$($p_x$ and $p_y$ are degenerated), $\mathrm{E_{2u}}$($f_{xyz}$ and $f_{(x^2-y^2)z}$ are degenerated), $\mathrm{B_{1u}}$($f_1$), $\mathrm{B_{2u}}$($f_2$) for spin-triplet state. 

In this section, we investigate $\lambda_{\mathrm{max}}$ using 3D band structure from beginning. Calculated Fermi surface in the case of $t_2 = 0.3$, $t_3 = -0.5$, $n = 0.45$ and $t_z = 1.4$ is shown in Fig.10. This well describes the band 37-electron. Such a low electron filling ($n = 0.45$) is due to the absence of electron in $k_z = \pm \pi$ plane. In Fig.11, we show the $U$-dependence of positive eigenvalues at $T = 0.01$. Because we consider the region far from the half-filling, the $d$-wave pairing states are suppressed by the large vertex corrections from large $U$. On the other hand, the $p$-wave pairings are stabilized by them. The two $p$-wave pairing ($\mathrm{A_{1u}}$ and $\mathrm{E_{1u}}$) are nearly degenerate, and the stable state between them is changed by the small variations of some other parameters such as $t_2$, $t_3$, $n$ and $t_z$. Thus it is difficult for us to determine which $p$-wave pairing is stable. Here, we note that the band width in this parameter equals 16.0, which is much larger than that in 2D calculations, thus such large value of $U$ in Fig.11 is usable. Next, in Fig.12, we show the $t_z$-dependence of eigenvalues for $p_x(\mathrm{E_{1u}})$-wave and $p_z(\mathrm{A_{1u}})$-wave pairing states at $U = 9$ and $T = 0.01$. The eigenvalue for $p_x(\mathrm{E_{1u}})$-wave pairing is suppressed by 3D dispersion $t_z$ as shown in \S 5.1. On the other hand, $p_z(\mathrm{A_{1u}})$-wave pairing is stabilized by $t_z$. These two $p$-wave states are nearly degenerate in the best fitting parameter region to $\mathrm{UPt_3}$ ($1.3 \le t_z \le 1.5$). The $p_z(\mathrm{A_{1u}})$-wave pairing agrees with experimental results which confirm the gap structure with a line of nodes at $k_z = 0$ plane. However, we cannot explain the $\mathrm{E_{2u}}$ hybrid gap ($f_{xyz}$ and $f_{(x^2-y^2)z}$). We may have to use more realistic models such as a multi-orbital model to explain it. 

\begin{figure}[t]
\begin{center}
\includegraphics[width=8cm]{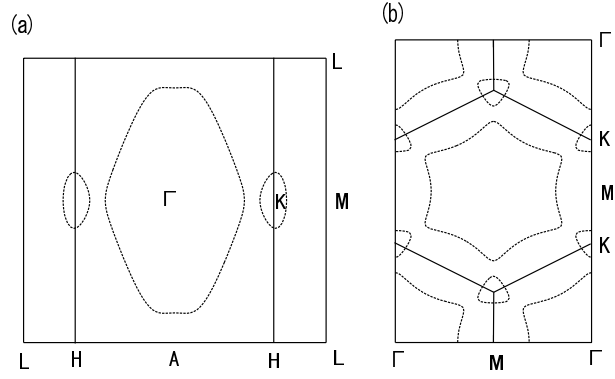}
\end{center}
\caption{The Fermi surface at $t_2 = 0.3$, $t_3 = -0.5$, $n = 0.45$ and $t_z = 1.4$ (a)in $k_y = 0$ plane, (b)in $k_z = 0$ plane.  This well describes the band 37-electron in ref.19.}
\label{f10}
\end{figure}

\begin{figure}[t]
\begin{center}
\includegraphics[width=8cm]{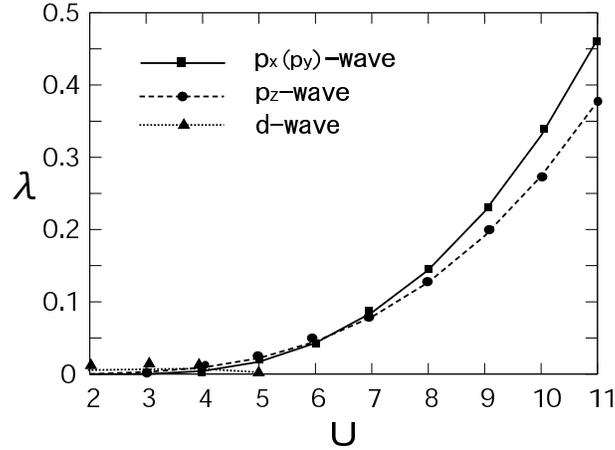}
\end{center}
\caption{The $U$-dependence of eigenvalues for $p_z$-wave ($\mathrm{A_{1u}}$) pairing and $p_x$($p_y$)-wave ($\mathrm{E_{1u}}$) pairing in the case of $t_2 = 0.3$, $t_3 = -0.5$, $n = 0.45$, $t_z = 1.4$ and $T = 0.01$. The band width is 16.0. The solid (dashed) line with black squares (circles) is the result for $p_x$-wave ($p_z$-wave) in TOPT, and the dotted line with black triangles is the result for $d_{xy}$($d_{x^2 - y^2}$)-wave ($\mathrm{E_{2g}}$). In large $U$ region, two $p$-wave pairings are stabilized by the vertex corrections, and these are nearly degenerate. }
\label{f11}
\end{figure}

\begin{figure}[t]
\begin{center}
\includegraphics[width=8cm]{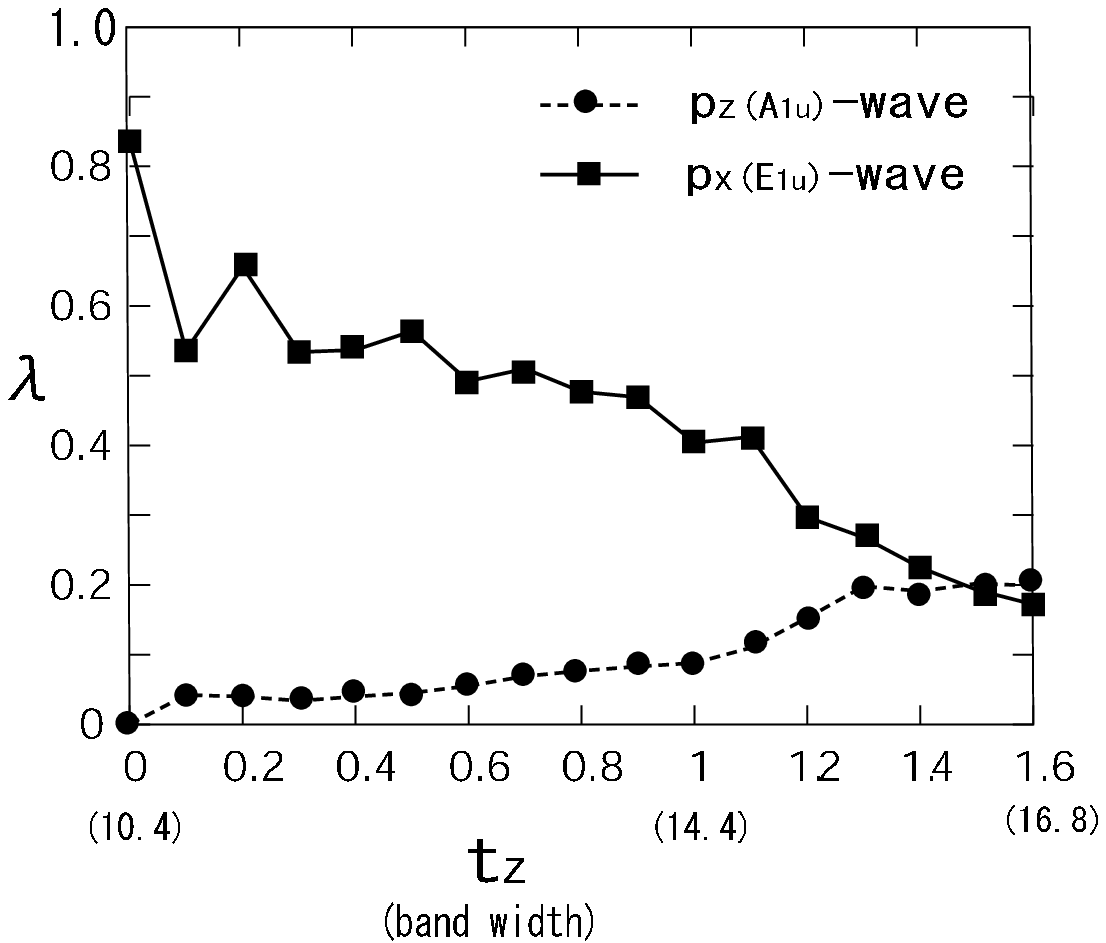}
\end{center}
\caption{The $t_z$-dependence of eigenvalues for $p_z$-wave $(\mathrm{A_{1u}})$ pairing and $p_x$-wave $(\mathrm{E_{1u}})$ pairing in the case of $t_2 = 0.3$, $t_3 = -0.5$,$n = 0.45$, $U = 9$ and $T = 0.01$. The solid (dashed) line with black squares (circles) is the result for $p_x$-wave ($p_z$-wave) in TOPT. The $p_z$-wave pairing stabilized by $t_z$ is nearly degenerate with the suppressed $p_x$(or $p_y$)-wave in the region $1.3 \le t_z \le 1.5$.}
\label{f11}
\end{figure}

\section{Discussions}
In this paper, we have investigated the single band Hubbard model for $\mathrm{UPt_3}$ on the basis of the perturbation theory. We have performed all the calculations in renormalized scheme, where we consider $\xi(\vec{k})$ to be the dispersion of quasiparticles with heavy effective mass. In that scheme, we should not include the normal selfenergy corrections giving rise to mass enhancement which has been already included by the renormalization. Therefore we have ignored them. As a result, we have also ignored the damping effect due to the normal selfenergy, but qualitative properties of superconductivity such as the pairing symmetry is not usually influenced\cite{yanase}. 

When we examine $T_{\mathrm{c}}$ quantitatively, we have to estimate the band width of the heavy quasiparticles. This is very difficult especially in the case of the heavy fermion systems. The effective mass and the band width contribute directly to the value of $T_{\mathrm{c}}$, as Sasaki $et \ al.$ account for the difference of $T_{\mathrm{c}}$ for high-$T_{\mathrm{c}}$ cuprates of $\mathrm{La}$ system and that of $\mathrm{Y}$ system using the $d-p$ model\cite{sasaki}. Thus we have to adopt more realistic models such as the periodic Anderson model and calculate the higher order selfenergy corrections, to examine both the heavy effective mass and $T_{\mathrm{c}}$ quantitatively.

Yamagami also calculated the energy band structure for $\mathrm{UPt_3}$\cite{yamagami} using a fully-relativistic spin-polarized version of the linearized augmented-plane-wave (LAPW) method. The calculated energy band structure show us fact that, compered with the one we calculated in subsection 5.2, the six pockets are large and the main rugby ball like Fermi surface locates on the flat bottom of the dispersion. Thus, we have introduced further neighbor hopping integrals to reproduce his energy band structure and then calculated eigenvalues more precisely. However, the calculated density of state at the Fermi surface and the calculated $T_{\mathrm{c}}$ do not distinctly differ from those in subsection 5.2. Therfore, we do not miss the essence of band 37-electron using the simple Fermi surface calculated in subsection 5.2. 

Our simple calculations predict the following points.
\begin{enumerate}
\item The $d$-wave pairing is not possible because the antiferromagnetic fluctuation is suppressed.
\item Only the triplt (odd parity) pairing is probable. 
\item Simple ferromagnetic fluctuation machanism can not explain the superconductivity of $\mathrm{UPt_3}$. This is because the van Hove singularities and the small pocket structures exist only near the $k_z = 0$ plane.
\item The $p$-wave pairing is the most probable. It is stabilized by the large vertex corrections from large $U$.
\item The calculated $T_{\mathrm{c}}$ for $p$-wave pairing is rather low. In the renormalized scheme, the limit of the effective on-site repulsion is not clear because of the ambiguity of the band width of the quasiparticles. In some case, we need to use larger value to realize the superconductivity of $\mathrm{UPt_3}$.
\end{enumerate}

\section{Conclusion}
In conclusion, on the basis of the microscopic theory, we have investigated the pairing symmetry and the transition temperture in the two-dimensional(2D) and three-dimensional(3D) hexagonal Hubbard model considering the band 37-electron in $\mathrm{UPt_3}$. We have solved the $\mathrm{\acute{E}liashberg}$ equation using the third order perturbation theory with respect to $U$. As the results of the 2D calculation, we obtain distinct two types of stable spin-triplet pairing states. One is the $f$-wave($\mathrm{B_1}$) pairing in large filling $n$ and small $U$ region, which is caused by the ferromagnetic fluctuation. Then the other is the $p_x$(or $p_y$)-wave($\mathrm{E_1}$) pairing in large $U$ region far from the half filling which is caused by the vertex corrections only. However, we find that the former $f$-wave pairing is easily destroyed by introduced 3D dispersion. This is because the 3D dispersion breaks the favorable structures for the $f$-wave pairing such as the van Hove singularities or the small pocket structures. Thus, the ferromagnetic fluctuation mediated spin-triplet state can not explain the superconductivity of $\mathrm{UPt_3}$. We have also studied the case of pairing symmetry with a polar gap. This $p_z$-wave($\mathrm{A_1}$) is stabilized by both the large hopping integral along c-axis $t_z$ and large $U$. It is nearly degenerate with the suppressed $p_x$(or $p_y$)-wave($\mathrm{E_1}$) in the best fitting parameter region to $\mathrm{UPt_3}$ ($1.3 \le t_z \le 1.5$). The $p_z$-wave pairing partially agree with the experimental results. Finally, we note that these two $p$-wave pairing states exist far from the half-filling, in which the vertex correction terms play crucial roles like the case in $\mathrm{Sr_2RuO_4}$.

\section*{Acknowledgements}
The auther S.S. is grateful to Dr. H. Ikeda and S. Sasaki for valuable discussions. Numerical calculation in this work was carried out at the Yukawa Institute Computer Facility.

\end{document}